**Highlights**

1. Dielectric unusual behavior has been successfully explained by the Rezlescu model.
2. Long τ (ns) is determined, can be utilized for memory and spintronics devices.
3. $a_{th}$ is calculated and well compared with $a_{expt}$.
4. Sn substituted Ni-Zn single phase inverse cubic spinel has been synthesized.

# Structural and electrical properties of Sn substituted double sintering derived Ni-Zn ferrite

M.A. Ali[a], M.N.I. Khan[b], F.-U.-Z. Chowdhury[a], S.M. Haque[b], M.M. Uddin[a,*]

[a] Department of Physics, Chittagong University of Engineering and Technology (CUET), Chittagong-4349, Bangladesh.
[b] Materials Science Division, Atomic Energy Center, Dhaka-1000, Bangladesh.

**Abstract**

The Sn substituted Ni-Zn ferrites were synthesized by the standard double sintering technique using nano powders of nickel oxide (NiO), zinc oxide (ZnO), iron oxide ($Fe_2O_3$) and tin oxide ($SnO_2$). The structural and electrical properties have been investigated by the X-ray diffraction, scanning electron microscopy, DC resistivity and dielectric measurements. Extra intermediate phase has been detected along with the inverse cubic spinel phase of Ni-Zn ferrite. Enhancement of grain size is observed in Sn substituted Ni-Zn ferrites. DC resistivity as a function of temperature has been investigated by two probe method. The DC resistivity was found to decrease whereas the dielectric constants increase with increasing Sn content in Ni-Zn ferrites. The dielectric constant of the as prepared samples is high enough to use these materials in miniaturized memory devices based capacitive components or energy storage principles.

[*] Corresponding author.
E-mail address: mohi@cuet.ac.bd (M. M. Uddin)

## 1. Introduction

In recent years, the spinel ferrites belong to $AB_2O_4$ structure having tetrahedral *A* site and octahedral B site have drawn huge attention due to their characteristic properties to meet the necessities in various applications. Remarkable progresses have been observed to invent and development of new ferrites. The research and application of magnetic materials have been developed considerably in the few past decades. The Ni-Zn ferrites have been found to be the most versatile ferrites systems from the viewpoint of their technological application because of its high electrical resistivity, high permeability, chemical stability and low eddy current losses [1-5], especially ideal for high frequency applications. The properties of Ni-Zn ferrites can be altered by changing chemical composition, preparation methods, sintering temperature ($T_s$) and impurity element or levels. The improvement of the basic properties of Ni-Zn ferrites regarding various applications have been reported [6-20] by altering chemical composition, doping ions or levels having different valence states. The tetravalent ions such as $Ti^{4+}$, $Sn^{4+}$ and $Si^{4+}$ substitution have greatly influenced the properties Ni-Zn ferrites [21].

Details investigation on $Ti^{4+}$ doping in Ni-Zn ferrite system has been carried out [6-8, 12, 17] while introduction of $Sn^{4+}$ has attracted less attention [7, 9]. Though some studies have mainly focused on the magnetic properties of Sn substituted Ni-Zn ferrites, the nonmagnetic properties such as electrical conductivity and dielectric properties are not reported. The materials with high-dielectric constants ($\geq 10^3$) have become immense interest for the miniaturized memory devices that are based on the capacitive components or energy storage principles [22, 23]. Moreover, investigations are limited in substitution of non-magnetic ions of $Fe^{3+}$ in Ni-Zn ferrite system. Simultaneous change of Ni and Zn by Sn substitution in the Ni-Zn ferrite system is essential to elucidate basic understanding and mechanism.

In this study, we have reported the structural and electrical properties of pure and Sn substituted Ni-Zn ferrite. To the best of our knowledge; this is the first detailed study on tin substituted Ni-Zn ferrite prepared by double sintering technique.

## 2. Materials and methods

Solid state reaction route was followed to synthesize Sn substituted Ni-Zn ferrite, $Ni_{0.6-x/2}Zn_{0.4-x/2}Sn_xFe_2O_4$ (x = 0.00, 0.05, 0.10, 0.15, 0.20 and 0.30) (NZSFO). We have used high purity (99.5%) (US Research Nanomaterials, Inc.) oxide precursors. The nano powders are taken as raw materials. The particle size of nickel oxide (NiO), zinc oxide (ZnO), iron oxide ($Fe_2O_3$) and tin oxide ($SnO_2$) are 20-40, 15-35, 35-45 and 35-55 nm, respectively. The preparation technique is described elsewhere [5]. The final sintering of the samples was carried out at 1300°C for 4 h in air and natural cooling was followed. Structural characterization of the synthesized samples was carried out by X-ray diffraction (XRD) using Philips X'pert Pro X-ray diffractometer (PW3040) with Cu-$K_\alpha$ radiation ($\lambda$ = 1.5405 Å) and scanning electron microscope (SEM). DC resistivity was measured using Keithley-6514 DC measurement system. Dielectric measurements were done by a Wayne Kerr precision impedance analyzer (6500B) in the frequency range of 10 Hz to 100 MHz with drive voltage 0.5V at room temperature.

## 3. Results and discussion

*3.1 Structural properties*

The XRD patterns of Sn substituted Ni-Zn ferrites with the chemical composition $Ni_{0.6-x/2}Zn_{0.4-x/2}Sn_xFe_2O_4$ (NZSFO) are shown in Fig. 1. It is seen that the observed peaks (111), (220), (311), (400), (422), (511), (440) and (533) confirmed the spinel structure of the $Ni_{0.6}Zn_{0.4}Fe_2O_4$ (NZFO) for x=0.0. The extra new intermediate phase of $NiSnO_3$ and $SnO_2$ is observed at around $2\theta=33.3°$ for the Sn concentration higher than that of $x > 0.1$. Similar extra phase of $NiSnO_3$ has also been observed and reported in Sn substituted NiFe$_2$O$_4$ ferrite system [24]. The intensity of extra phase increases with the increase of Sn concentration. The corresponding positions of all the sharp peaks were used to obtain the interplanar spacing. The lattice parameter for each peak of the samples was calculated using the equation $a = d\sqrt{h^2 + k^2 + l^2}$.

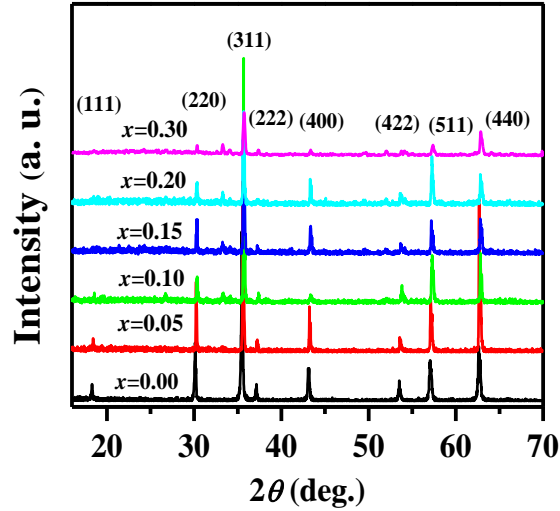

Fig. 1. The X-ray diffraction patterns of NZSFO (x = 0.0, 0.05, 0.1, 0.15, 0.2 and 0.3) ferrites samples.

To determine the exact lattice parameter, Nelson–Riley (N-R) extrapolation method was used. The N-R function is represented by the equation $F(\theta) = \frac{1}{2}\left[\frac{\cos^2\theta}{\sin^2\theta} + \frac{\cos^2\theta}{\theta}\right]^{1/2}$. The exact lattice parameter a$_0$ was determined using least square fit method from the plot of lattice parameter 'a' of each peak versus F(θ) [figure not shown].

Theoretical calculation of lattice parameter can also be done using the following equation $a_{th} = \frac{8}{3\sqrt{3}}\left[(r_A + R_0) + \sqrt{3}(r_B + R_0)\right]$, where R$_0$ is the radius of the oxygen ion (1.32 Å) [3] and r$_A$ and r$_B$ are the ionic radii of the tetrahedral (A-site) and octahedral (B-site) sites, respectively [25]. The values of r$_A$ and r$_B$ can be calculated from the cation distribution of the system and be represented by

$r_A = C_{AFe}r(Fe^{3+}) + C_{AZn}r(Cd^{2+}) + C_{ASn}r(Sn^{4+})$ and $r_B = \frac{1}{2}[C_{BNi}r(Ni^{2+}) + C_{BFe}r(Fe^{3+}) + C_{BSn}r(Sn^{4+})]$ [26, 27].

The information of cation distribution can be used to know about the magnetic behavior of ferrite sample. The materials with desired properties for practical application can be developed with the help of cation distribution [28]. The cation distribution is assumed based on the hypothesis that Sn has a tendency to occupy tetrahedral (A) site at lower concentration, whereas it occupies the octahedral site at higher concentration. The cation distribution of A and B sites for each substitution level (Sn content) is presented in Table 1. The ionic radii for Fe, Ni, Zn and Sn are 0.65, 0.69, 0.75 and 0.69 Å, respectively.

The effect of Sn substitution on the lattice constant, $a_{expt}$ is shown in Fig. 2 (a). It is found that the lattice constant initially decreases up to $x = 0.1$ and thereafter it increases at $x = 0.15$, again it decreases up to $x = 0.3$ and finally increases for x > 0.3. It indicates that the variation of $a$ with $x$ does not obey the Vegard's law [29]. Our experimental results follow nonlinear trend with $x$ which is consistent with the reported observation for Sn substituted $NiFe_2O_4$ [24]. The variation of theoretical lattice constant with Sn content is also shown in Fig. 2 (a). The similar trend for both experimental and calculated lattice constant is observed. The lattice constant of all the doped composition is less than that of the parent one. A decreasing trend in lattice constant with an increase in the content of Sn can be attributed to the ionic size differences since the unit cell has to contract when substituted by ions with smaller size. The ionic radius of the $Sn^{4+}$ and $Zn^{2+}$ is 0.69 and 0.75Å, respectively. The partial replacement of $Zn^{2+}$ by $Sn^{4+}$ might be expected to cause shrinkage of the unit cell. It can be noted that the ionic radii of Sn and Ni is same (0.69Å), hence the substitution of Sn for Ni does not affect the lattice constant value.

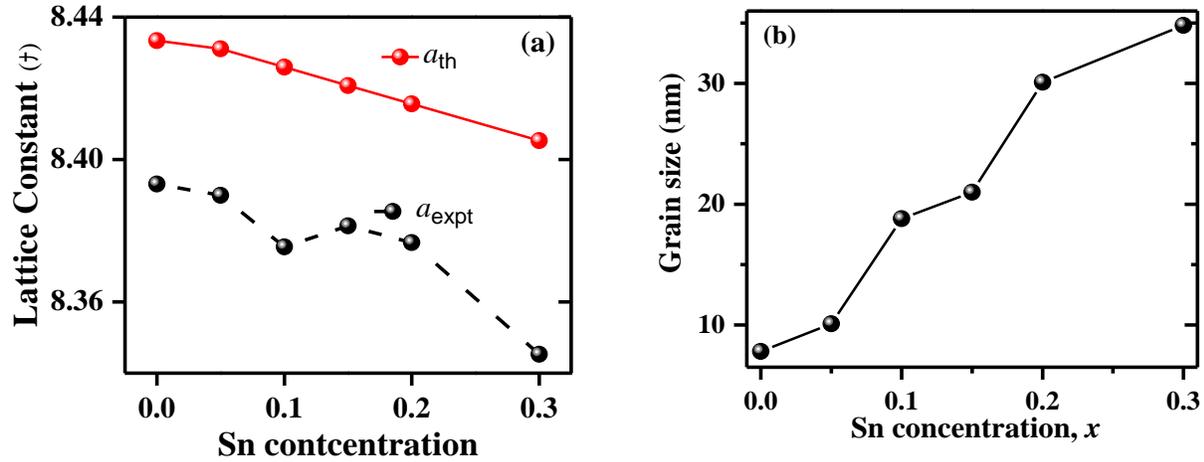

Fig. 2. (a) The experimental and theoretical lattice constants; (b) the average grain size as a function of Sn concentration (x = 0.0, 0.05, 0.1, 0.15, 0.2 and 0.3) of NZSFO ferrites.

The X-ray density ($\rho_{x\text{-ray}}$), bulk density ($\rho_b$) and porosity (P) of the NZSFO ferrites are presented in Table 1. Normally the $\rho_b$ of the same composition is smaller than the $\rho_{x\text{-ray}}$. This can be explained by the existence of pores within the samples which are developing during the sintering process and depend on the sintering temperatures, conditions and time. The $\rho_b$ of doped sample (NZSFO) is less than that of the parent (NZFO). The porosity of the NZSFO increases almost linearly with Sn doping concentration and relatively higher values are observed. The porosity in the samples is strongly dependent on the amount of applied pressure during sample preparation. In the present case, the applied pressure is 10 kN/cm$^2$ (~1 ton/cm$^2$).

*3.2 Microstructure study*

Fig. 3 (a-g) shows the SEM micrographs of Sn substituted Ni-Zn ferrite taken at room temperature. Clear grains and grain boundaries are evident from the micrographs. SEM micrographs reveal the polycrystalline nature of microstructures with grains of different shapes and size. The linear intercept technique has been used to calculate the average grain size (grain diameter) and values are given in Table 1 for different Sn concentration [30]. Micrographs show that the grains are almost homogenously distributed throughout the sample surface.

**Table 1**
Variations of lattice parameter, X-ray density, bulk density, average grain size, porosity and activation energy of (NZSFO).

| Sn content, x | Chemical formula | A site | B site | $r_A$ (Å) | $r_B$ (Å) | $a_{th}$ (Å) | $a_{exp}$ (Å) | $\rho_{x-ray}$ (gm/cc) | $\rho_b$ $d_B$ (gm/cc) | $D_g$ (μm) | P (%) | $E_a$ (eV) |
|---|---|---|---|---|---|---|---|---|---|---|---|---|
| 0.0 | $Ni_{0.6}Zn_{0.4}Fe_2O_4$ | $FeZn_{0.4}$ | $FeNi_{0.6}$ | 0.95 | 0.532 | 8.43341 | 8.39311 | 5.32 | 4.28 | 07.8 | 19.6 | 0.19 |
| 0.05 | $Ni_{.575}Zn_{.375}Sn_{.05}Fe_2O_4$ | $FeZn_{0.375}Sn_{0.025}$ | $[FeNi_{0.575}Sn_{0.025}]O_4^{2-}$ | 0.948 | 0.532 | 8.4311 | 8.38996 | 5.41 | 3.73 | 10.1 | 31.0 | 0.119 |
| 0.1 | $Ni_{.55}Zn_{.35}Sn_{.1}Fe_2O_4$ | $FeZn_{0.35}Sn_{0.03}$ | $[FeNi_{0.55}Sn_{0.07}]O_4^{2-}$ | 0.933 | 0.538 | 8.42595 | 8.37546 | 5.52 | 3.85 | 18.8 | 30.2 | 0.116 |
| 0.15 | $Ni_{.525}Zn_{.325}Sn_{.15}Fe_2O_4$ | $FeZn_{0.325}Sn_{0.035}$ | $[FeNi_{0.525}Sn_{0.0115}]O_4^{2}$ | 0.917 | 0.545 | 8.42079 | 8.38137 | 5.59 | 4.10 | 21.0 | 26.6 | 0.1023 |
| 0.2 | $Ni_{.5}Zn_{.3}Sn_{.2}Fe_2O_4$ | $FeZn_{0.3}Sn_{0.04}$ | $[FeNi_{0.5}Sn_{0.016}]O_4^{2-}$ | 0.902 | 0.552 | 8.41563 | 8.37665 | 5.68 | 3.80 | 30.1 | 33.1 | 0.099 |
| 0.3 | $Ni_{.45}Zn_{.25}Sn_{.3}Fe_2O_4$ | $FeZn_{0.25}Sn_{0.05}$ | $[FeNi_{0.45}Sn_{0.25}]O_4^{2-}$ | 0.872 | 0.566 | 8.40532 | 8.34531 | 5.91 | 3.87 | 34.8 | 34.5 | 0.11 |

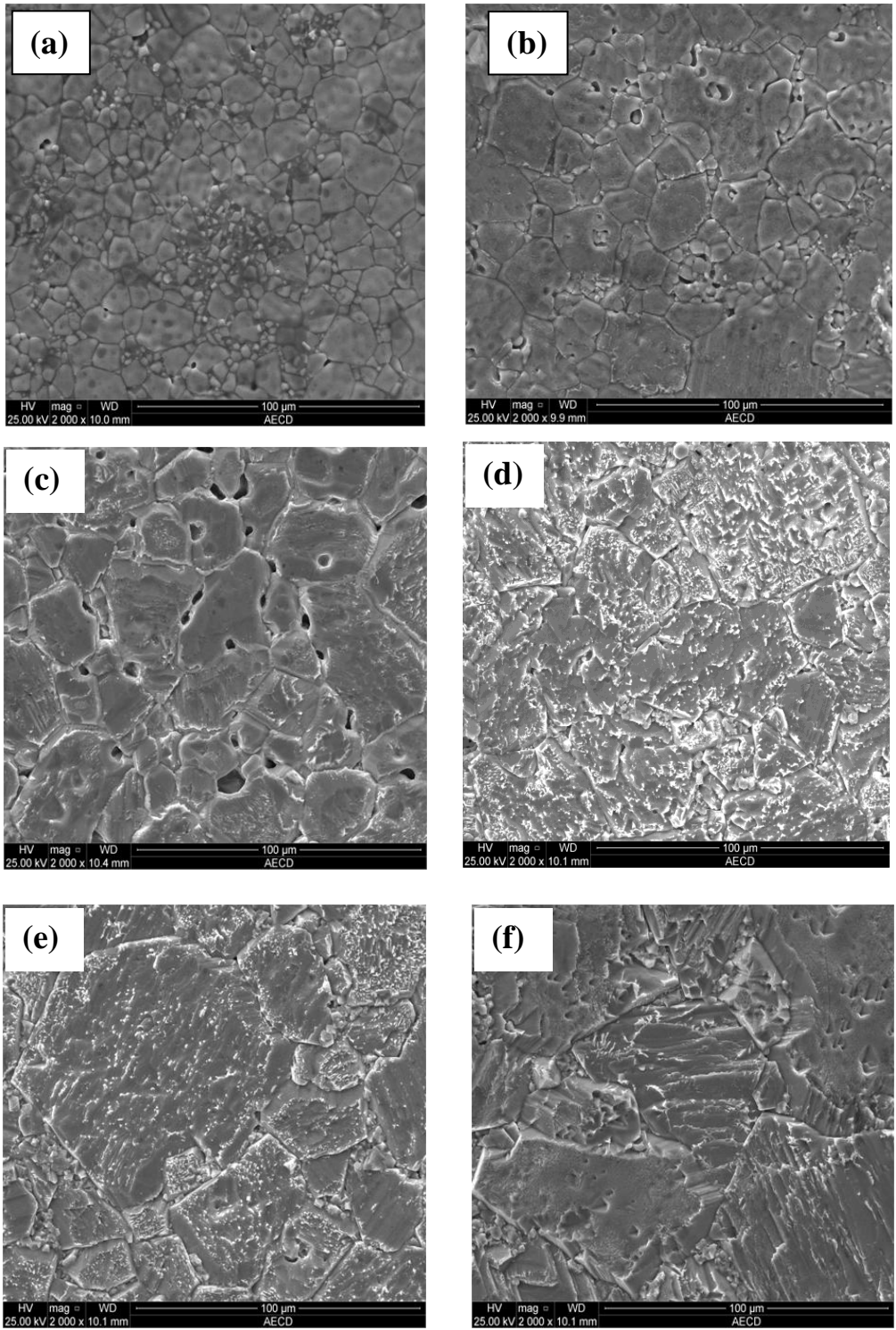

Fig. 3. SEM micrographs of the NZSFO ferrite for (a) $x = 0.0$, (b) $x = 0.05$, (c) $x = 0.1$, (d) $x = 0.15$, (e) $x = 0.2$, and (f) $x = 0.3$.

It is seen from the Fig. 2 (b) that the grain size increases with increasing Sn content. Some oxides, like $SnO_2$, could bring down the melting point. So for the same sintering temeprature, presence of Sn helps to sinter better or it is equivalent to raising the sintering temperature for the ferrite. As we know that the grain size is increased with increasing sintering temperature. As a result enhancement of grain might be expected.

### 3.3. Electrical properties

*3.3.1. DC resistivity*

DC resistivity of the NZSFO samples was measured by two probe method and is plotted as a function of temperature in Fig. 4 (a). It is observed that the resistivity decreases exponentially with increasing temperature indicating semiconductor behavior of the prepared ferrites. It is also found that the resistivity decreases with increasing Sn contents x which can be explained as a consequence of microstructural and structural modification owing to the change in composition. It is observed from the SEM micrographs that the grain size increases with increasing Sn concentration. As a result the number of grains and the grain boundaries decreases. The insulating behavior of grain boundaries may be attributed to the bulk of the resistivity in ferrite [31]. The grain size increases faster than the porosity in NZSFO with the increase of Sn contents. The average grain size for x = 0.0 (NZFO) is found to be around 4.2-7.8 μm, while for the NZSFO ferrite the range is 10-34 μm. Though the porosity causes an increase in resistivity but the increase in grain size results in a decrease in resistivity. The combined effect of the two events might decrease the resisitivty of prepared ferrites. In addition, another reason is that Sn simultaneously substituted Zn ions (prefer tetrahedral A site) and Ni ions (prefer octahedral B site) [32], this will lead to the migration of $Fe^{3+}$ and $Fe^{2+}$ (which is responsible for electric conduction in ferrite) and the fluctuation of valence states for tin as $Sn^{2+}$ and $Sn^{4+}$ increases the electronic exchange resulting the resistivity decreases. The reduction of resistivity might be a result of the combined effect.

The DC electrical resistivity as a function temperature of the samples can be presented by the Arrhenius type equation: $\rho = \rho_0 exp\left(E_a/2KT\right)$, where $\rho$ and $\rho_0$ are resistivity of samples at any temperature and room temperature, respectively. The parameter $E_a$ is the activation energy; k is the Boltzmann constant (= $8.62 \times 10^{-5}$ eV) and T is the absolute temperature. The activation energy in the ferromagnetic region of the sample was calculated from the plot of Fig. 4 (b) using the relation $E_a = slope \times 4.606 \times 8.62 \times 10^{-5}$ eV [33] and presented in Table 1.

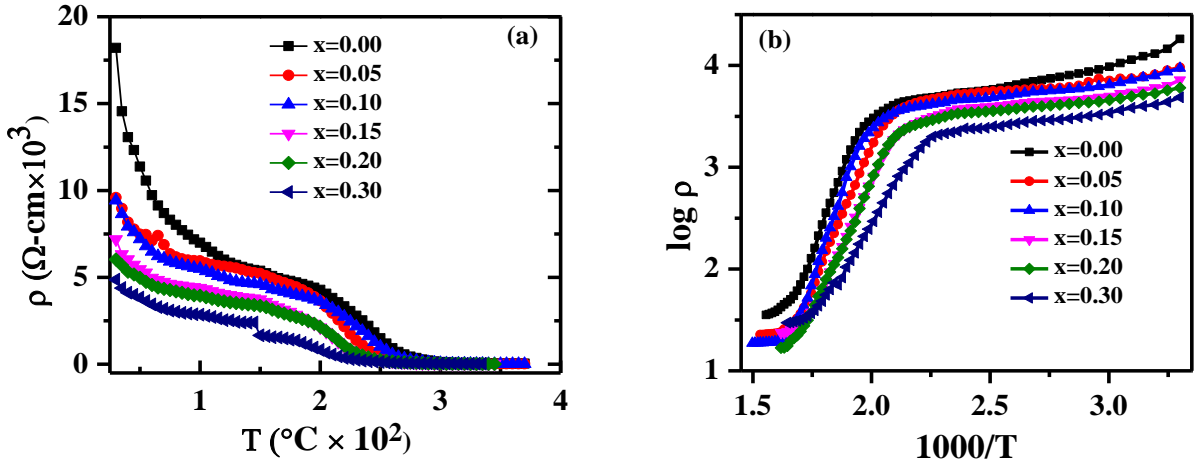

Fig. 4. (a) Variation of resistivity as a function of temperature and (b) log$\rho$ vs. 1000/T graph for different Sn concentration.

The value of $E_a$ is found to be decreased with increasing Sn content which confirms the electronic character of the conduction process. The value of activation energy is higher for the sample with high electrical resistivity. Our calculated values are consistent with this conclusion except for $x = 0.3$ [34]. The conduction process due to electron exchange between $Fe^{3+}$ and $Fe^{2+}$ may be accelerated with the increase of Sn content.

*3.3.2. Dielectric relaxation properties*

The dielectric permittivity ($\varepsilon'$) [$\varepsilon' = CL/\varepsilon_0 A$, where, C is the capacitance, L is the thickness, A is the cross-sectional area of the flat surface of the pellet and $\varepsilon_0$ is the constant of permittivity for free space] and tanδ of the NZSFO ceramics for ($x$ = 0.0, 0.05, 0.1, 0.15, 0.2 and 0.3) as a function of frequency are illustrated in Fig. 5. At lower frequency the value of $\varepsilon'$ is much higher while it shows very small value at high frequency. It is also observed that the $\varepsilon'$ shows fast decreasing trend with increasing frequency at lower frequencies, whereas it decreases slowly at higher frequencies even becomes almost zero and independent of frequency. This can be elucidated by the Koop's phenomenological theory based on the Maxwell-Wagner model [35-37] considering inhomogeneous double layer dielectric structure. The Koop's theory assumes that the ferrites are composed of well conducting grains separated by a thin layer of poorly conducting grain boundaries. The grains with high conductivity are formed during the ferrite preparation. These grains are separated by poorly conducting grain boundaries. These grain boundaries could be formed during the sintering process due to the superficial reduction or oxidation of crystallites in the porous materials as a result of their direct contact with the firing atmosphere [38]. The poorly conducting grain boundaries have been found to be effective at lower frequencies while ferrite fairly conducting grains is effective at high frequencies [39]. Therefore, the values of $\varepsilon'$ are found to be higher at lower frequencies and with increasing frequency it decreases.

It is found that the dielectric constant of Sn substituted NZSFO ferrite is higher than that of NZFO. The observed variation in the dielectric constant with Sn concentration could be explained on the basis of local displacement of charge carriers in presence of external electric field and octahedral (B) site occupancy of Sn ions. In NZSFO ferrite system, two probable conduction mechanisms, viz. electron hopping between $Fe^{3+}$ and $Fe^{2+}$ and hole hopping between $Ni^{3+}$ and $Ni^{2+}$ ions might be operative. Ferrite system containing Ni when sintered and cooled in air, a considerable amount of oxygen is absorbed, giving rise to the formation of $Ni^{3+}$ ions. In oxygen rich region, conduction takes place through $Ni^{2+} \leftrightarrow Ni^{3+}$ and in oxygen deficient regions, conduction takes place through electron hopping between $Fe^{3+} \leftrightarrow Fe^{2+}$. Tin cations ($Sn^{2+}$ and $Sn^{4+}$) have greater tendency to occupy the octahedral sites in comparison to tetrahedral sites [40]. Since Sn is replacing nickel at B site (with the increase of Sn concentrations), a large number of $Fe^{3+}$ ions are expected to present in B site and there is a possibility of electron exchange between $Fe^{2+} \leftrightarrow Fe^{3+}$ due to Zn volatilization. This can also be attributed from the fluctuation of valence states of $Sn^{2+}$ and $Sn^{4+}$. Moreover, microstructure has great influence in dielectric constant. The dielectric constant of ferrite generally increases with increasing grain sizes. Large grain size has noticeable difference between grain and grain boundaries resistances which enhances polarization and hence dielectric constant [41].

Frequency dependent loss factor (tanδ, is defined as $\varepsilon''/\varepsilon'$) for different Sn concentration of NZSFO is shown in Fig. 5 (b). The curves show the dielectric relaxation peaks at a particular frequency [42]. The dielectric relaxation peaks appear when the externally applied AC electric field becomes equal to that of the jumping frequency of localized electric charge carrier [43]. The unusual behavior of the dielectric in ferrites could be successfully explained by the Rezlescu model which states that the collective contribution of both types of electric charge carriers (electron and hole) to the dielectric polarization is the main source for that type of dielectric relaxation [44]. In ferrites, the electrical conduction arises due to the electron exchange between $Fe^{2+}$ and $Fe^{3+}$ and hole transfer between $Ni^{3+}$ and $Ni^{2+}$ at the octahedral (B) sites which is similar to that of dielectric polarization in ferrites [44, 45]. The ions of Fe and Ni are formed by the following mechanism: $Ni^{2+} + Fe^{3+} \leftrightarrow Ni^{3+} + Fe^{2+}$ and $Fe^{3+} \leftrightarrow Fe^{2+} + e^{-}$.

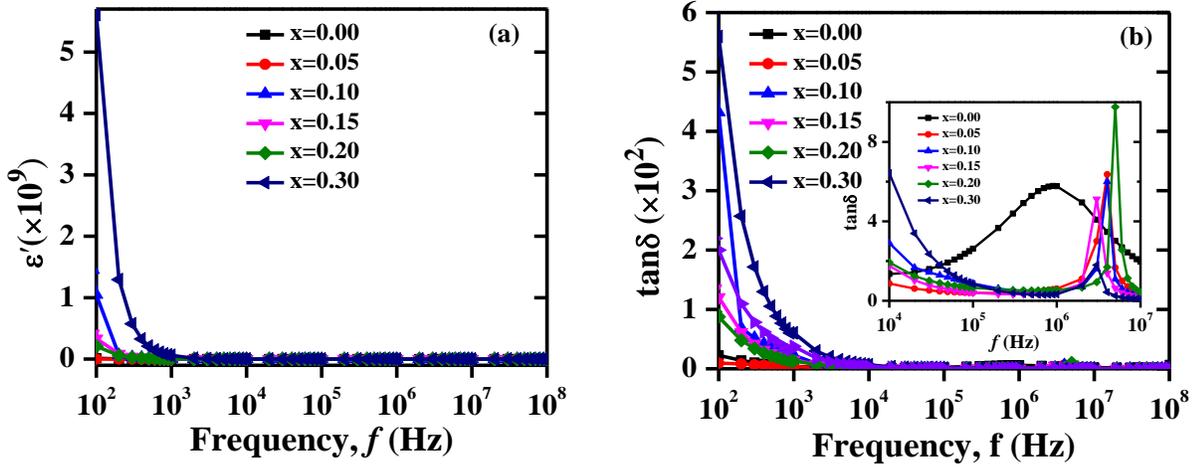

Fig. 5. Frequency dependence of (a) dielectric constants and (b) dielectric loss tangent for different Sn concentration. [Inset: formation of peaks].

*3.3.3. Electric modulus*

By the study of complex dielectric modulus the information about the electrical response of the materials can be understood. It also gives information on the nature of polycrystalline samples (homogenous or inhomogeneous). The electric modulus provides ideas about the electrical relaxation process of a conducting material [46]. In order to reconfirm the relaxation process in the sample the real and imaginary parts of the electric modulus are calculated using the relations

$$M^* = 1/\varepsilon^*$$

$$M'' = \frac{\varepsilon''}{(\varepsilon'^2 + \varepsilon''^2)}$$

$$M' = \frac{\varepsilon'}{(\varepsilon'^2 + \varepsilon''^2)}$$

It is observed that the value of $M'(\omega)$ is very low (approaching zero) in the low frequency region and continuously increasing with the rise in frequency by showing a tendency to saturate at maximum asymptotic value (i.e., $M_\infty = 1/e_\infty$) (Fig. 6), which indicates the short range mobility of the charge carrier conduction process in the samples [47].

The variation of $M''(\omega)$ as a function of frequency for different Sn concentration is characterized by a clearly resolved peak in the pattern. Significant asymmetry in the peak with their positions lying in the dispersion region of $M'(\omega)$ and $M''(\omega)$ versus frequency pattern is observed. The low frequency side of the $M''(\omega)$ peak determines the range in which charge carriers can move over long distances i.e., successful hopping of charge carriers is possible. The high frequency side of the $M''(\omega)$ peak determines the range in which the charge carriers are spatially confined to their potential wells and being mobile over short distances. Thus, the peak frequency is indicative of transition from long range to short range mobility with increase in frequency. The peak frequency shifts towards higher value with increasing Sn contents. The characteristic frequency at which $M''(\omega)$ is maximum ($M''_{max}$) corresponds to relaxation frequency and is used for the evaluation of relaxation time, $\tau_{M''}[= 1/(2\pi f_{M''})]$. The dielectric relaxation time ($\tau$) is found to be 8 ns and 159 ns for $x= 0.0$ and 0.05 and ~80 ns for $0.05<x<0.3$.

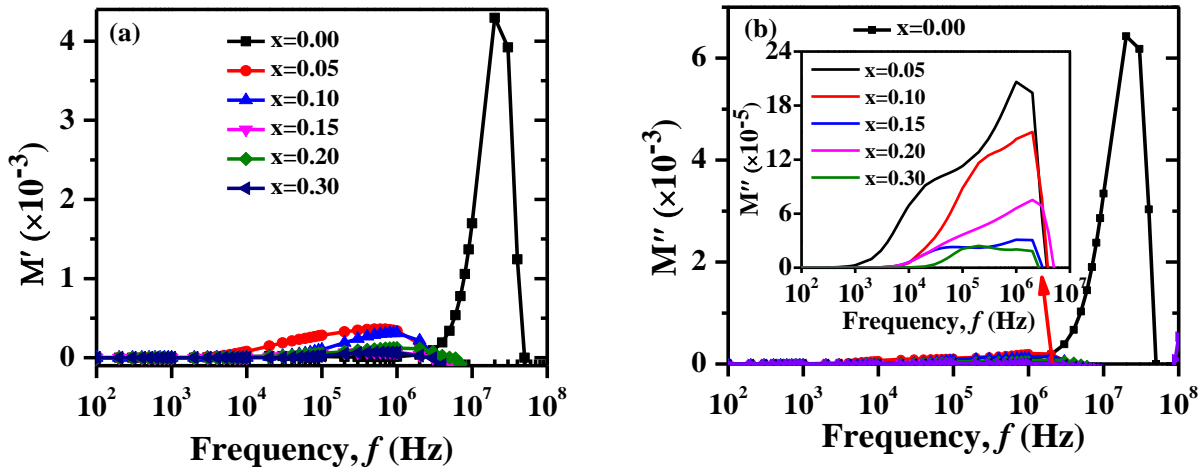

Fig. 6. Variation of (a) real and (b) imaginary part of electric modulus of NZSFO.

### 3.3.4. Impedance spectroscopy

A complete perception of the electrical properties of the electro-ceramic materials such as impedance of electrodes, grain and grain boundaries can be understood by the complex impedance spectrum technique (Cole-Cole plot or Nyquist plot). A Cole-Cole plot typically consists of two successive semicircles: first semicircle is due to the contribution of the grain boundary at low frequency and second is due to the grain or bulk properties at high frequency of the materials.

Fig. 7 (a) shows the variation of real part of impedance as a function of frequency for different doping concentration of NZSFO. The pattern of variation of Z´ shows dispersion in the low frequency region followed by a small plateau and, finally, all the curves coalesce. This behavior indicates an increase in conduction with doping and frequency. At the higher frequencies, the impedance (Z´) curves are merged for the samples ($x = 0.0 – 0.3$) shows a possible release of space charge results the space charge polarization is reduced.

Fig. 7 (b) illustrates the frequency response of the imaginary part of impedance for the NZSFO. It can be seen that the maxima value of Z″ shifts towards higher frequency side for up to $x = 0.1$ while it shifts slightly lower frequency for further increasing Sn up to $x = 0.2$. Furthermore, at $x = 0$, two peaks are observed, which represent the combined contribution of the grain and grain boundary.

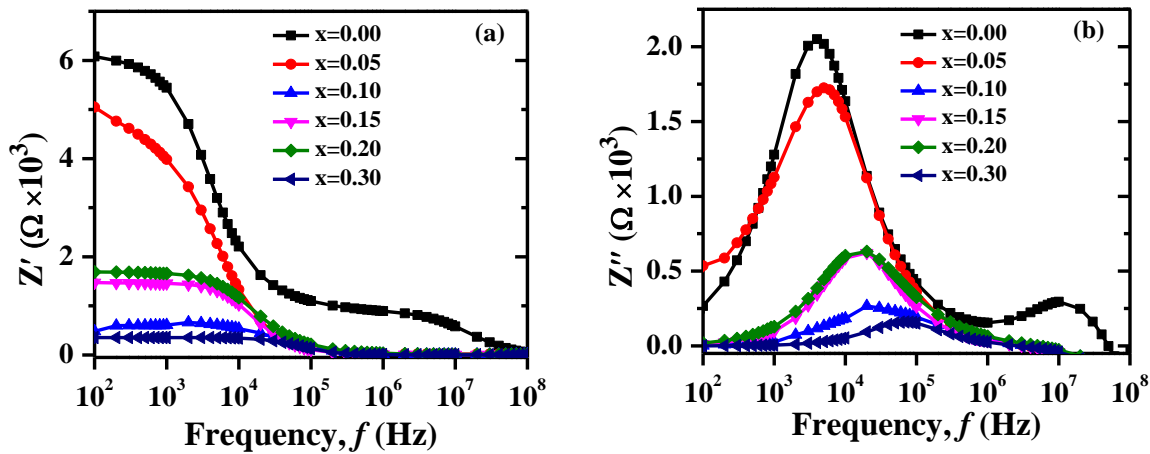

Fig. 7. Variation of (a) real and (b) imaginary part of complex impedance of NZSFO.

The complex impedance spectra (Cole-Cole plot) of the sample measured for different Sn concentration are shown in Fig. 8(a). The Cole-Cole plot typically shows two partially overlapping semicircular arcs at low temperatures with center lying slightly below the real axis suggesting the departure from ideal Debye type of relaxation process. Two very clear semicircular arcs have been observed for x = 0.0 for NZFO, whereas only one semicircular arc is observed for NZSFO (x = 0.05, 0.1, 0.15, 0.2 and 0.3). The presence of one of the semicircular arcs is diminished with the introduction of Sn concentration. The observed two overlapping semicircular arcs for pure NZFO are due to the contribution of the grain (bulk) and grain boundary to electrical properties of the material [5]. The contribution of grains is dominant in the NZSFO samples leading to a single semicircular arc. It can be understood that grain size (diameter) increases with increasing Sn contents in the NZSFO [shown in Fig. 2 (b)] which means grain boundary decreases. The contribution of grain boundary is also decreased results one semicircle in the NZSFO samples. Fig. 8 (b) represents the RC equivalent circuit for single semicircle for the NZSFO samples.

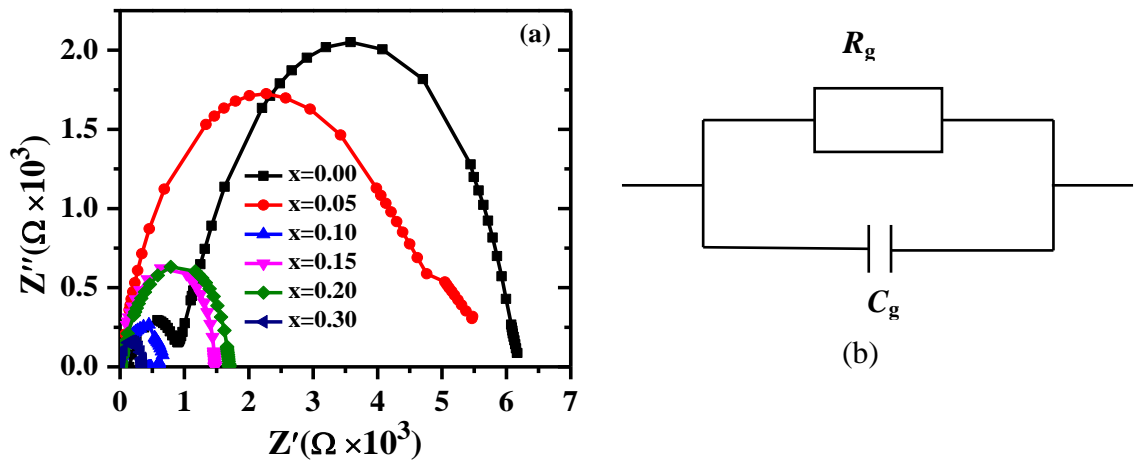

Fig. 8. (a) Cole-Cole plot of the NZSFO (b) equivalent model circuit for single semicircle Cole-Cole plot.

## 4. Conclusions

The NZSFO (x = 0.0, 0.05, 0.1, 0.15, 0.2 and 0.3) ferrites, sintered at 1300 °C, have been synthesized and the structural and electrical properties have been studied. The presence of Sn ions has a remarkable influence on the properties of Ni-Zn ferrites. The lattice constant is found to be decreased whereas enhancement of grain size is found in Sn substituted Ni-Zn ferrites. A decrease in resistivity, i.e. enhancement of conductivity and dielectrics constant, is also noticed for NZSFO (x = 0.05, 0.1, 0.15, 0.2 and 0.3). The unusual behavior of the dielectric in ferrites could be successfully explained by the Rezlescu model. Irregular long range and non-Debye type dielectric relaxation is observed in the NZSFO. The complex impedance spectra show principal contribution of grain resistance for the NZSFO, confirmed by the classical Cole-Cole complex impedance spectra. Long spin dielectric relaxation times in several nano second ranges have been observed in the NZSFO. This long spin relaxation time makes NZSFO as one of the promising candidates for future memory materials and spintronics device applications.

**Acknowledgment**

The authors are grateful to the Directorate of Research and Extension, CUET for arranging the financial support for this work.

**Table Caption**

Table 1 Variations of lattice parameter, X-ray density, bulk density, average grain size, porosity and activation energy of (NZSFO).

**Figure Captions**

Fig. 1. The X-ray diffraction patterns of NZSFO (x = 0.0, 0.05, 0.1, 0.15, 0.2 and 0.3) ferrites samples.

Fig. 2. (a) The experimental and theoretical lattice constants; (b) the average grain size as a function of Sn concentration (x = 0.0, 0.05, 0.1, 0.15, 0.2 and 0.3) of NZSFO ferrites.

Fig. 3. SEM micrographs of the NZSFO ferrite for (a) x = 0.0, (b) x = 0.05, (c) x = 0.1, (d) x = 0.15, (e) x = 0.2, and (f) x = 0.3.

Fig. 4. (a) Variation of resistivity as a function of temperature and (b) log$\rho$ vs. 1000/T graph for different Sn concentration.

Fig. 5. Frequency dependence of (a) dielectric constants and (b) dielectric loss tangent for different Sn concentration. [Inset: formation of peaks].

Fig. 6. Variation of (a) real and (b) imaginary part of electric modulus of NZSFO.

Fig. 7. Variation of (a) real and (b) imaginary part of complex impedance of NZSFO.

Fig. 8. (a) Cole-Cole plot of the NZSFO (b) equivalent model circuit for single semicircle Cole-Cole plot.